# On proximity detection systems for pico-projectors
Dr. Edward Buckley

## Abstract

*Proximity detection systems have been proposed as a potentially beneficial method for increasing the eye-safe luminous flux of laser-based picoprojectors. In this letter it is shown that, whilst the benefit for panel-based systems could be significant, the impact upon scanned-beam projectors is far smaller.*

## 1. Introduction

"Pico-projector" products, which are typically marketed as small battery-powered devices consuming less than 5 W capable of providing a luminous flux of 10–20 lm, began to emerge in 2008 and were initially based on LED light sources. Lasers could potentially offer a number of advantages over LEDs, with associated system advantages including a small form-factor, long depth of field, polarization independence and potentially higher efficiencies.

To date, a number of laser light-engine architectures have been proposed and demonstrate. Lasers can be used as light sources for conventional imaging architectures, illuminating a small amplitude-modulating liquid-crystal– on–silicon (LCOS) panel with small projection optics used to magnify the resultant field. Scanned-beam projectors represent an alternative approach, in which a rapidly moving silicon micromirror is employed to mechanically deflect a rapidly modulated laser spot across the image.

Since the publication of recent laser safety analyses for pico-projectors [2, 3], the use of proximity detection systems has been considered as a potential solution to the luminous flux limitations imposed on laser projection systems by Class 1 and Class 2 laser safety classifications. Although such a technique is not specifically mentioned in the current IEC 60825-1 standard [1], it has been suggested that a proximity detection technique could allow the projector to output a luminous flux above the previously-determined eye-safe limits by enabling automatic shut-off should an obstruction occur at a measurement distance of less than $r$ from the projector aperture.

Since current laser safety standards effectively impose a maximum luminous flux $L_{max}$ at a distance of $r = 100$ mm, but the eye-safe radiometric power increases with $r$, a higher $L_{max}$ for $r > 100$ mm could be achieved in principle if a proximity detector were employed. In this letter, we investigate the impact of $r$ on the maximum radiometric power $P_{max}$, and hence the luminous flux as a functiosn of $r$, $L_{max}(r)$, that can be achieved with the use of proximity detection in scanned-beam and panel-based laser projectors.

## 2. Scanned-beam laser projectors

### 2.1. Class 2

In a Class 2 analysis [3] the measurement distance $r$ affects three things; the first is the acceptance angle $\gamma$, defined such that

$$\tan\left(\frac{\gamma}{2}\right) = \frac{d}{2r} \qquad (1)$$

where $r$ is the measurement distance and $d$ is the diameter of the eye. For $d/r \ll 1$, which is the case in this analysis, then we have

$$\gamma \approx \frac{d}{r} \qquad (2)$$

so the measurement distance $r$ is inversely proportional to acceptance angle $\gamma$. We also know that the proportion of output luminous flux delivered to the eye $\eta$ is

$$\eta \approx \frac{\gamma^2}{\theta_h \theta_v} \qquad (3)$$

where $\theta_h$ and $\theta_v$ are the horizontal and vertical projection angles respectively. It follows that the maximum Class 2 power $P_{max}$ is proportional to $\gamma^{-2}$ and hence $P_{max} \sim r^2$, where $\sim$ is used to mean "varies as." An illustration of these parameters is provided in Figure 1 below.

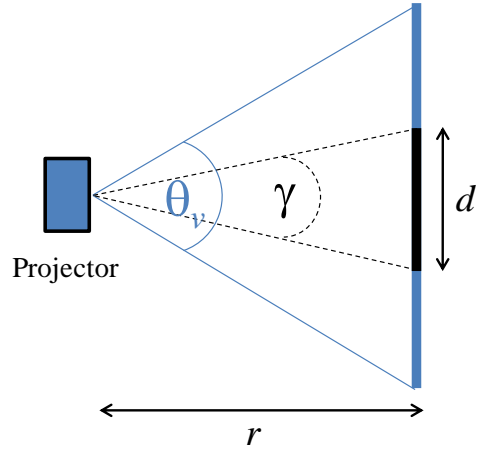

**Figure 1** –Measurement geometry for projection systems considered in this study.

Next is the number of pulses delivered to the eye, $n$. For a scanned-beam projection system in which $N$ is the vertical resolution and $f_r$ is the frame rate, then the number of pulses incident upon the measurement aperture $n$ is

$$n = \frac{N\gamma}{\theta_v} f_r T_2 \qquad (4)$$

and $T_2 = 0.25$ s is the classification period in Class 2. It is clear that the number of pulses $n$ is proportional to the acceptance angle $\gamma$. From equation (20) in [3] then for a given pulse duration $T_i$, the accessible exposure limit (AEL)

$$AEL = 7\times10^{-4} C_6 \frac{(nT_i)^{0.75}}{T_2} \text{ W} \qquad (5)$$

and, since $\gamma$ is inversely proportional to $r$ from equation (2) and because the the maximum radiometric power $P_{max}$ is directly proportional to the AEL, we have that the power $P_{max} \sim r^{-3/4}$.

Finally, the angular extent of the source $\alpha$ is also related to $r$. For a scanned-beam system which forms $N$ scan lines each containing spots of size $d_{spot}$, then the source angular extent $\alpha$ is given by

$$\tan\left(\frac{\alpha}{2}\right) = \frac{d_{spot}}{2r} \qquad (6)$$

and for $d_{spot} / r \ll 1$,

$$\alpha \approx \frac{d_{spot}}{r} \qquad (7)$$

so it is clear that the source angular extent is inversely proportional to $r$. A schematic of the scan pattern intercepted by the measurement aperture of diameter $d$ is shown in Figure 2 below.

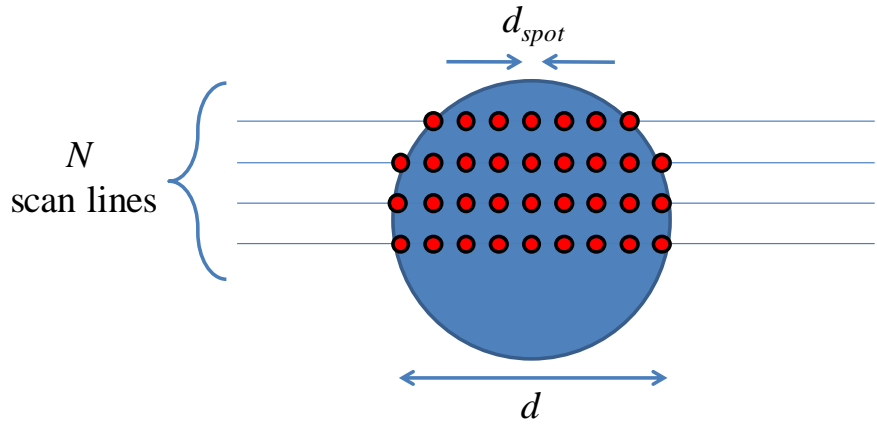

**Figure 2** – $N$ scan lines intercepted by measurement aperture of diameter $d$.

The angular extent is linearly related to the effective source size correction factor $C_6$ given by

$$C_6 = \begin{cases} 1 & \alpha \leq \alpha_{min} \\ \alpha/\alpha_{min} & \alpha_{min} \leq \alpha \leq \alpha_{max} \\ \alpha_{max}/\alpha_{min} & \alpha > \alpha_{max} \end{cases} \qquad (8)$$

where $\alpha_{min}$ = 1.5 mrad and $\alpha_{max}$ = 100 mrad, and for a scanned-beam projector we have two cases to consider. In the region $1.5 \leq \alpha \leq 100$, we know from Freeman et al. [4] that $d_{spot} \leq 0.6$ mm and hence $6 \text{ mm} \leq r \leq 400$ mm. Using equation (8) again, it follows that $P_{max} \sim r^{-1}$ in this region. For $r \geq 400$ mm, the optical design of a scanned-beam projector ensures that $d_{spot}$ is proportional to $r$; this gives $\alpha \leq 1$ mrad.so that $C_6 = 1$ and $P_{max} \sim$ const.

In summary, then, we have that the maximum radiometric power depends upon three terms; $r^2$, since only a fraction of the radiation is delivered to the eye, $r^{-1}$ or a constant, depending upon the source angular extent determined by the distance at which the radiation is measured, and $r^{-3/4}$ due to the fact that multiple scan lines intercept the measurement aperture. It follows that either $P_{max} \sim r^2 \times r^{-1} \times r^{-3/4}$ or $P_{max} \sim r^2 \times \text{const} \times r^{-3/4}$ and so if the maximum luminous flux at $r = 100$ mm is $L_{max}(r = 100 \text{ mm})$ then the luminous flux as a function of $r \geq 100$ mm, $L_{max}(r \geq 100 \text{ mm})$, is

$$L_{max}(r \geq 100 \text{mm}) = L_{max}(r = 100 \text{mm}) \cdot \begin{cases} \left(\dfrac{r}{100}\right)^{1/4} & r \leq 400 \text{ mm} \\ \sqrt{2}\left(\dfrac{r}{400}\right)^{5/4} & r > 400 \text{ mm} \end{cases} \quad (9)$$

where the factor of $\sqrt{2}$ is due to the breakpoint at $r = 400$ mm. We can calculate the Class 2 luminous flux gain $G$ compared to the $r = 100$mm case as

$$G = \frac{L_{max}(r \geq 100 \text{mm})}{L_{max}(r = 100 \text{mm})} \quad (10)$$

This relation demonstrates quite clearly the limited use of a proximity sensor for increasing Class 2 luminous flux. If the measurement distance $r$ is doubled to $r = 200$ mm, then the total maximum Class 2 radiometric power $P_{max}$, and hence photometric power $L_{max}$, only increases by a factor of $2^{1/4} = 20\%$. Beyond $r = 400$ mm the situation is improved, although for a luminous flux gain of a factor of two, the proximity detector would need to be set for an observation distance of 528 mm. To achieve luminous flux levels similar to those provided by Class 1 LCOS-based projectors [5] would require a four-fold increase in luminous flux with $r = 920$ mm and it is debatable whether this scenario is consistent with proposed handheld pico-projector use cases.

## 2.2. Class 1

The Class 1 photochemical power limit for the blue and green wavelengths is given by [1]

$$P_{b,g} = \left(\frac{\alpha}{11}\right)^2 \frac{AEL_{b,g}}{\eta t} \text{ W} \quad (11)$$

where $AEL_{b,g}$ is the acceptable exposure limit (AEL) at the blue and green wavelengths and $t = 100$ s. It is clear that the radiometric power depends only upon the acceptance angle $\gamma$ (due to $\eta$ in equation (3)) and the source subtense $\alpha$ and, since $\eta \sim r^{-2}$ and $\alpha \sim r^{-1}$, then it follows that $P_{b,g}$ (and hence $L_{max}$) is independent of $r$. It is therefore not possible to realize a Class 1 luminous flux gain by increasing the measurement distance $r$.

## 3. Panel-based projectors

### 3.1. Class 2

According to the analysis in [2], the Class 2 eye-safe radiometric output power for a panel-based projector depends upon only the acceptance angle $\gamma$ and source angular extent $\alpha$. Since it is fixed by geometry, the dependence upon the acceptance angle is the same as for the scanned-beam case so that $P_{max} \sim r^2$.

In a panel-based projector employing a diffuser in the projection lens telescope, the angular extent of the source is determined by

$$\alpha = \frac{2f}{r} \tan\left(\frac{\theta}{2}\right) \qquad (12)$$

where $f$ is the focal length of the projection lens and $\theta$ the diffuser scatter angle [2]. So we have $\alpha \sim r^{-1}$ and from equation (8) it follows that $P_{max} \sim r^{-1}$. The maximum radiometric power that can be delivered by a panel-based projectors is therefore governed by two terms; $r^2$, because only a fraction of the radiation is captured by the measurement aperture, and $r^{-1}$ due to the angular extent of the source. So we have that $P_{max} \sim r^2 \times r^{-1}$ i.e. $L_{max} \sim P_{max} \sim r$ and, since $\alpha > \alpha_{min}$ for a large range of $r$ and for sensible values of $f$ and $\theta$, it is reasonable to suppose that for this projection architecture the luminous flux gain is linearly related to the measurement distance beyond $r = 100$ mm. The Class 2 luminous flux gains as a function of the measurement distance $r$ for scanned-beam and panel-based projectors are plotted in
Figure 3 below.

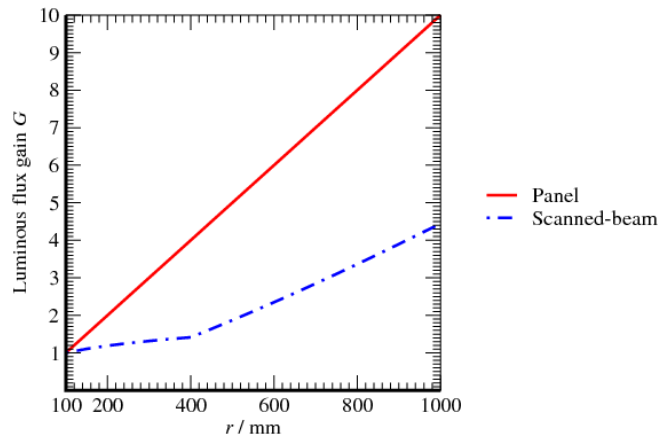

**Figure 3** – Luminous flux gain $G$ as a function of $r$ for scanned-beam and panel projectors.

Since LCOS panel projectors are theoretically already capable of delivering several hundred lumens in Class 2, proximity-detection systems could be of real value in achieving high levels of brightness for situations in which the projector is stationary. This scenario could be well-suited to laser projectors used for office or digital cinema-type applications.

### 3.2. Class 1

The analysis is the same as for the scanned-beam case; Class 1 luminous flux is independent of measurement distance.

### 4. Summary

Proximity detection systems have been proposed as a method of increasing the measurement distance $r$ beyond the $r = 100$ mm limit prescribed by IEC 60825-1, thereby potentially allowing higher eye-safe luminous flux values.

For scanned-beam projectors, the increase in Class 2 luminous flux using this method would be small since the maximum Class 2 luminous flux only scales as $r^{1/4}$ for $r \leq 400$ mm and as $r^{5/4}$ beyond $r = 400$ mm. For panel-based projectors the benefit could be significantly greater, since the maximum Class 2 luminous flux scales as $r$ for a wide range of $r$. In both cases, the Class 1 luminous flux is independent of $r$.

## List of Figures